# Electrical characterization of fully encapsulated ultra thin black phosphorus-based heterostructures with graphene contacts

*Ahmet Avsar[1], Ivan J. Vera-Marun[1,2], Tan Jun You[1], Kenji Watanabe[3], Takashi Taniguchi[3] Antonio H. Castro Neto[1] and Barbaros Özyilmaz[1,4,†]*

[1]Graphene Research Center & Department of Physics, 2 Science Drive 3, National University of Singapore, Singapore 17542, Singapore

[2]Physics of Nanodevices, Zernike Institute for Advanced Materials, University of Groningen, Nijenborgh 4, 9747 AG, Groningen, The Netherlands

[3]National Institute for Materials Science, 1-1 Namiki, Tsukuba 305-0044, Japan

[4]Nanocore, 4 Engineering Drive 3, National University of Singapore, Singapore 117576, Singapore

[†]Corresponding author: barbaros@nus.edu.sg,

The presence of finite bandgap and high mobility in semiconductor few-layer black phosphorus offers an attractive prospect for using this material in future two-dimensional electronic devices. Here we demonstrate for the first time fully encapsulated ultrathin (down to bilayer) black phosphorus field effect transistors in Van der Waals heterostructures to preclude their stability and degradation problems which have limited their potential for applications. Introducing monolayer graphene in our device architecture for one-atom-thick conformal source-drain electrodes enables a chemically inert boron nitride dielectric to tightly seal the black phosphorus surface. This architecture, generally applicable for other sensitive two-dimensional crystals, results in stable transport characteristics which are hysteresis free and identical both under high vacuum and ambient conditions. Remarkably, our graphene electrodes lead to contacts not dominated by thermionic emission, solving the issue of Schottky barrier limited transport in the technologically relevant two-terminal field effect transistor geometry.

**KEYWORDS** Black phosphorus, graphene, work-function, Schottky barrier, encapsulation, hysteresis

Monolayer black phosphorus (BP) or phopshorene, a one atom thick sheet of phosphorus atoms arranged in a puckered structure, has recently emerged as a potential candidate to address the shortcomings of previously studied 2D materials [1–4]. Few-layer BP can be easily cleaved from bulk crystals due to weak interlayer van der Waals interaction. Its transport properties depend strongly on the thickness of the obtained crystals and it has a direct band gap, increasing monotonically from ~0.3 eV in bulk to ~ 1.7 eV in phosphorene[5]. These properties, together with recent observations of high mobilities up to 1,000 cm$^2$/V.s[1,3,4] at room temperature (one order of magnitude higher than previous semiconducting 2Ds[6]) make BP suitable for applications such as fast and broadband photodetectors[7], solar cells[8] and digital electronics[1].

The cornerstone of digital electronics is the field effect transistor (FET), which consists of an active semiconductor where charge carriers flow between source and drain electrodes and an electrostatic gate modulating the carrier density in the channel. For practical applications, the source and drain electrodes are expected to exhibit low contact resistances and linear responses, while the semiconducting channel should be stable over time. FETs using BP channel exhibited high on/off ratios[1,2,4,9], low subthreshold swings[1,10] and excellent current saturations[1,4] making BP attractive for the device applications. However the preparation of BP-based devices for such FET applications still poses challenges. The BP channel tends to degrade after a short exposure to air due to a photo-activated charge transfer process[11]. This degradation causes an increase in surface roughness and presence of chemisorped species leading to gap states in BP[12]. The resulting non-ideal interfaces, together with the presence of large Schottky barriers common in the physics of metal/semiconductor contacts set limitations for the performance of BP FETs. When transport across the contacts is dominated by the thermionic emission due to large Schottky barrir height, rather than tunneling, the charge injection efficiency is severely limited due to large contact resistances and a highly non-linear response.

We offer a complete solution to the challenges mentioned above by creating a 2D heterostructure where all interfaces are based on van der Waals bonding[13]. Towards this, we have contacted atomically thin layers of BP with monolayer graphene and fully encapsulated the device with hexagonal boron nitride (h-BN) in a layer-by-layer fashion, all under an inert argon environment with negligible water and oxygen contents. Introducing graphene as a conformal, one-atom-thick crystal, allows the ultrathin h-BN crystal to fully conform to the underlying structure without leaving any space for air or other species to diffuse. It is worth noting that utilizing regularly used metal contacts rather than monolayer graphene does not allow a full sealing even they are encapsulated with h-BN layer (See Supplementary Information). Such a perfect encapsulation leads to the observation of equally good performances under both ambient and vacuum conditions with hysteresis free characteristics in our pristine heterostructures. Integration of graphene electrodes with tunable work-function[14] enables enhanced charge injection efficiency into BP through low contact resistance interfaces with negligible Schottky barriers not dominated by thermionic emission but by tunneling essential for practical FET applications. Our results are important milestone towards fabrication of high quality BP devices relevant for operation under ambient conditions and room temperature.

The device fabrication starts with etching of mechanically exfoliated monolayer graphene flakes into isolated contact stripes by using standard electron beam lithography and oxygen plasma techniques. Few layers of h-BN crystal are partially deposited onto these monolayer graphene stripes with a dry transfer method[15,16]. The graphene/h-BN stack is isolated from the wafer with a KOH etching process. Meanwhile, a thin BP crystal is exfoliated onto $SiO_2$ (300nm)/Si wafer in an argon environment, followed by the transfer of the graphene/h-BN stack onto this freshly exfoliated BP crystal. We take care that h-BN fully encapsulates the BP crystal. During the device fabrication process, the surface of BP has never been exposed to air. Fabrication is completed with forming Cr/Au (2nm/80nm) electrodes on non-encapsulated regions of the graphene stripes and on top of the h-BN crystal, for serving as source-drain and top gate electrodes, respectively. A more detailed description of the device fabrication is discussed in the supplementary information.

Optical images of the device before and after the metallization process are shown in Figure1-a and Figure 1-b, respectively. The final device is annealed at 340 C under high vacuum conditions. This annealing step is crucial to obtain high quality contacts. Firstly, the annealing treatment improves the bonding of graphene and h-BN layers to the BP crystal by removing the small bubbles that might be formed during the transfer process[15]. Secondly, it removes the possible adsorbants present at the interface of BP and $SiO_2$ that can cause significant hysteresis in transport measurements[17]. AFM images confirm that the device is clean and free of wrinkles (See Supplementary Information). The resulting heterostructure is illustrated schematically in Figure 1-c. Raman spectrum for the encapsulated BP device is shown in Figure 1-d where the characteristic[18] $A_g^1$, $B_{2g}$ and $A_g^2$ peaks of BP are clearly visible at the wave numbers of 362.6, 439.8 and 467.4, respectively. We do not observe any obvious change in Raman spectrum before and after annealing at 340 C, implying that the encapsulating h-BN crystal effectively protects BP. It is worth noting that non-encapsulated BP crystals are observed to degrade during the annealing process (See supplementary Information). Electronic transport measurements are carried out in two-terminal configuration under ambient and vacuum conditions (~ $1 \times 10^{-7}$ Torr). In this work, we studied a total of three samples and present representative data on two encapsulated BP devices with the thickness of ~ 4.5 nm (6 layer, sample S1) and ~ 1.6 nm (2 layer, sample S2) respectively. Unless otherwise stated, the results obtained in the relatively thicker sample are discussed in the manuscript. $SiO_2$ (~ 300 nm) and h-BN (~ 16 nm) dielectrics are utilized to apply back and top gate biases ($V_{BG}$ and $V_{TG}$) respectively (See Figure 2-a). The bias current ($I_{SD}$) of BP-based devices is characterized as a function of source-drain voltage ($V_{SD}$), temperature (T), $V_{BG}$ and $V_{TG}$.

Now we discuss the transport characteristics of our encapsulated device measured under both vacuum and ambient conditions. For comparison, we also fabricated a non-encapsulated device with similar thickness of BP crystal (Insets of Figure 2-b and 2-c). Figure 2-b shows the $V_{TG}$ dependence of $I_{SD}$ at fixed $V_{BG}$ (-40 V) and $V_{SD}$ (100 meV) under vacuum conditions. While the non-encapsulated device shows a significant hysteresis (~30V) even under vacuum conditions, the encapsulated device has nearly hysteresis free transport characteristics. In general, such positive hysteresis is caused by the sensitivity of 2D crystals to the external chemical environments at the top surface and with the hosting substrate (See supplementary information)[17,19]. In our device, the first is excluded as the device is fully encapsulated. The second potential source is also eliminated due to the high temperature annealing treatment during device fabrication. The stability of our device architecture is evidenced by noting that the device shows this hysteresis free transport behavior even two months after the device is fabricated (See supplementary information). Next, we discuss the transport performance under ambient conditions. As can be seen in Figure 2-c, the encapsulated device has nearly identical output characteristics under both ambient and vacuum conditions. On the other hand, for the non-encapsulated device we observe a significant drop in the conductivity (up to 80%) at ambient conditions (Inset Figure 2-c), resulting in lower field effect mobility and current modulation (See supplementary Information). These observations clearly indicate that the high quality h-BN encapsulating layer together with monolayer graphene electrodes protect the BP surface from interaction with air. Compared to other passivation schemes such as $SiO_2$, PMMA and ALD grown dielectrics of $Al_2O_3$ and $HfO_2$; h-BN is intrinsically inert and pinhole free[20,21]. This minimizes the possible oxidation of BP through pinholes in the dielectric and also any chemical reaction. It was recently shown that the water pecursor in ALD grown $Al_2O_3$ causes hysteresis in $MoS_2$-based FETs[19,22]. Such effect will be more dramatic in BP based devices as the crystal is very sensitive to the adsorbed oxygen in water.[23] On the contrary, the interface in our heterostructure device is pristine as demonstrated by the lack of hysteresis and the stability discussed above.

Now we turn our attention to the equally important contact resistance aspect of our device. Figure 3-a-c show the 2D color plots of $I_{SD}$ as a function of $V_{SD}$ and $V_{BG}$ at $V_{TG}$ = -4V, 0V, 4V respectively. The device shows p-type behavior even at $V_{TG}$= 4V. However, the application of $V_{TG}$ causes a clear shift in the threshold region which starts to be observed at more positive $V_{BG}$ values as $V_{TG}$ is changed from positive to negative values. Also note that $I_{SD}$ shows a tendency of saturation at high negative $V_{BG}$ values. To elucidate this behavior, we plotted the $V_{BG}$ dependence of $I_{SD}$ at $V_{SD}$ = 0.1V for $V_{TG}$ = -4V, 0V, 4V (Figure 3-d). While we see similar saturation behavior for each values of $V_{TG}$, the magnitude of the saturation current increases as the $V_{TG}$ is tuned from positive to negative values. It has been extensively discussed that the saturation current in 2D semiconductors-based FETs is dominated by the contact resistance in a two-terminal FET structure with only a minor contribution in our case of graphene and BP channels[4]. Therefore the observation of the increase in the saturation current as the $V_{TG}$ is changed from 4V to -4V clearly indicates that contact quality is improved. To quantify this, the total resistance vs $V_{BG}$ curve is extrapolated as suggested by R. Toy et al. by using the polynomial $A+B/x+C/x^2$ where the fitting parameter A refers to the contribution of the contact resistance to the total device resistance at high gate biases where the channel resistance is substantially small[24]. We observe a sharp decrease in the magnitude of the contact resistance from ~46 kΩ to ~ 15.8 kΩ by only changing the $V_{TG}$ from 4V to -4V. This is the same magnitude with the one obtained from the stable $MoS_2$-based heterostructure device which is similarly contacted and encapsulated with graphene and h-BN, respectively[24]. Such improvement is because of the unique work-function tunability of graphene electrodes which allows for a higher efficiency of hole injection to the valance band of BP at high electric fields.[14] Here, having $V_{TG}$ together with $V_{BG}$ in our device allows us to modulate the graphene work function more effectively compared to the one that can be achieved only with $V_{BG}$. As schematically shown in Figure 3-f, the application of large negative displacement fields (both $V_{TG}$ and $V_{BG}$ are negative) shifts the Fermi level of graphene towards the valence band of BP. The conduction and valence bands are bent upwards, making it easier to inject holes from graphene source-drain electrodes to BP. The relation between $I_{SD}$ and $V_{SD}$ gives additional information about the Schottky barrier heights. As shown in Figure 3-e, $I_{SD}$-$V_{SD}$ is slightly non-linear at small displacement fields ($V_{TG}$ = 4V, $V_{BG}$ = -30V). However, it becomes more linear as $V_{TG}$ is swept to 0V and eventually we observe a perfectly linear relation upon the application of large displacement fields ($V_{TG}$ = 4V, $V_{BG}$ = -30V). This observation gives independent proof for the reduced Schottky barrier at the source-drain interfaces. It is also worth noting that both encapsulated and non-encapsulated BP devices with Ti/Au contacts have non-linear $I_{SD}$-$V_{SD}$ behavior even at low $V_{SD}$ values in sharp contrast to this graphene contacted device (See supplementary information).

In order to determine the nature of the graphene/BP interfaces, we carried out a detailed study of $I_{SD}$ as a function of temperature. Figure 4-a shows the $V_{SD}$ and T dependence of $I_{SD}$ at large displacement fields ($V_{BG}$ = -70 V and $V_{TG}$ = -4 V). We observe the linear relation between $I_{SD}$ and $V_{SD}$ at both current polarities for all temperature. Remarkably, the change in $I_{SD}$ is less than 10% within the temperature range of 300 K to 10 K at high bias regime. This shows that thermionic emission is not the dominant transport mechanism in these graphene contacted BP devices. To confirm this, we analyze the data using the termionic emission model describing the charge transport through a schottky barrier into the BP channel

$$I_{SD} = AA^*T^{1.5}exp\left[-\frac{e}{k_BT}\left(\Phi_B - \frac{V_{SD}}{n}\right)\right]$$

where $A$ is the contact area, $A^*$ is the 2D Richardson constant, $e$ is the electron charge, $k_B$ is the Boltzmann constant and $\Phi_B$ is the Schottky barrier height and $n$ is the ideality factor[25]. Figure 4-b shows the Arrhenius plot (ln ($I_{SD}/T^{1.5}$) vs 1000/T) for source-drain bias voltages varying from 1 V to 0.1 V at fixed $V_{BG}$=-70 V and $V_{TG}$=-4 V. The Arrhenius plot has two distinct linear temperature regimes of high (60K-300K) and low (10K-60K) temperature. The intercepts ($S_0$) of slope vs $V_{SD}$ plot allow extracting the Schottky barrier height ($\Phi_B$) by using $\Phi_B = -10^3 S_0 k_B/q$ (See supplementary information).[25] All the extracted Schottky barrier height values are found to be small and negative, ranging from -17 meV to -2.5 meV. This indicates that thermionic emission does not dominate the transport in our devices. It is worth noting that these observations persist for each $V_{BG}$ values (See supplementary Information). As schematically illustrated in Figure 4-c, charge injection into semiconductors is dominated by tunneling (direct or thermally assisted) or thermionic emission. In contrast to a recent report on $TiO_2$/Co contacted BP device[10] which has charge injection dominated by thermionic emission, the nearly temperature insensitivity of $I_{SD}$ in our device suggests that charge injection is rather dominated by tunneling. Therefore our improved contacts should have significant impact on the transport in two terminal BP-based FETs. To demonstrate the latter, we show in Figure 4-d the temperature dependence of the mobility. The device demonstrates remarkable temperature stability. While a previous report on Ti/Au contacted BP device[4] has shown a decrease in mobility of more than 80% in a two-terminal measurement configuration as the temperature is reduced from RT to 20 K, our device shows a negligible decrease of mobility from RT down to 100 K (~10%) and only a moderate decrease down to 10 K ( ~30%). We note that temperature dependence of the mobility is even weaker at higher $V_{SD}$ values as discussed before and shown in inset of Figure 4-a. Observation of such a weak temperature dependence of the mobility is further demonstration of the negligible Schottky barrier height, and is potentially useful for achieving subthreshold swing below the kT/q thermal limit of 60 mV/decade at room temperature[26] in BP-based FETs if the dielectric is engineered (See supplementary information).

Finally we remark that our device architecture allows us to study very thin layers of BP, otherwise very challenging due to the issue of its rapid degradation. The recent prediction of band gap modulation in dual gated bilayer BP device[8] make our approach attractive for the perpective of fundamental science as well. Towards this, we utilized our encapsulated device architecture to study the transport properties of a bilayer BP-FET device. Figure 5 shows the $V_{BG}$ dependence of $I_{SD}$ at fixed $V_{SD}$ = 0.1 V. Similar to the previously shown device, bilayer BP-based FET also shows a dominant p-type behavior. However, we achieve a clear bipolar behavior where both electron and hole conduction are observed. This result is consistent with a strongly reduced electric field screening in such ultra thin samples[27]. In this device, we observe an on/off ratio of ~ 100 and a sub-threshold swing of ~30 V and ~ 17 V per decade at hole and electron conduction side respectively at room temperature. While this sub-threshold swing value is comparable to the previous results obtained with similar $SiO_2$ dielectric thicknesses[1,10], we note that improved sub-threshold swing values are obtained in thicker BP samples (See supplementary information). Field effect mobilities of ~ 120 $cm^2$/V.s at hole side and ~ 40 $cm^2$/V.s at electron side are extracted in this encapsulated bilayer BP device by using[1,2,4]

$$\mu = \frac{L}{W} \frac{1}{C_{OX}} \frac{1}{V_{SD}} \frac{\partial I_{SD}}{\partial V_{SD}}$$

where $\mu$ is the field effect mobility, $L$ and $W$ are the length and width of measured junction respectively, $C_{OX} = \frac{\varepsilon \varepsilon_{OX}}{d}$ is the capacitance of $SiO_2$ substrate. The enhanced gate tunability now allows us to tune the graphene work-function more effectively even with $V_{BG}$ alone and as a result, a linear $I_{DS}$-$V_{DS}$ relation is achieved even at high $V_{SD}$ values (Inset Figure 5). We note that unlike previously studied semiconducting 2D crystals[28,29], the encapsulation and enhanced graphene contacts do not improve the mobility of the encapsulated devices compared to non-encapsulated devices significantly. This suggests that the mobility is limited by the $SiO_2$ substrate rather than either the adsorbants present in air nor high Schottky barrier heights as previously suggested[1,4].

In conclusion, we report the fabrication of a new device architecture that allows charge transport studies in ultra thin BP crystals reliably by utilizing monolayer graphene and ultra thin h-BN as source-drain contacts and encapsulating layer. We demonstrate that these devices are extremely robust to the environment and hysteresis is nearly eliminated. Utilizing work-function tunable graphene electrodes into our heterostructure reduces the Schottky barrier height significantly, with current injection dominated by tunneling rather than thermionic emission. Our findings provide an important step toward fundamental and applied studies in few layers of BP-based FET as well as give access to study the transport properties of other sensetive crystals such as silicine[30] and GaSe[31].

## ACKNOWLEDGEMENTS

We thank J. Balakrishnan, E. C. T. O'Farrell, T. Taychatanapat and S. Natarajan for their help and useful discussions. B. Ö. would like to acknowledge support by the National Research Foundation, Prime Minister's Office, the Singapore under its Competitive Research Programme (CRP Award No. NRF-CRP9-2011-3), the Singapore National Research Foundation Fellowship award (RF2008-07), and the SMF-NUS Research Horizons Award 2009-Phase II. A.H.C.N. would like to acknowledge support by the National Research Foundation, Prime Minister's Office, Singapore under its Competitive Research Programme (CRP Award No. NRF-CRP6-2010-5). I. J. V. M. would like to acknowledge support by the Netherlands Organisation for Scientific Research (NWO).

**FIGURE CAPTIONS**

**Figure 1.** (a) Optical image of the device after the final transfer process. Red and black dashed areas show the black phosphorus crystal and one of the graphene stripes respectively. (b) Optical picture of the device after the contacts are formed. (c) The schematics of the the atomically sharp interfaces in encapsulated BP device. The black, gray, blue and yellow spheres represent the phosphorus, carbon, boron and nitrogen atoms respectively. (d) Raman spectroscopy of few layers of encapsulated black phosphorus flake before and after annealing process.

**Figure 2.** (a) The schematics of graphene contacted BP device together with the electrical connections for the device characterization. (b) Top gate voltage dependence of bias current in encapsulated device at fixed back gate (-40V) and source-drain (100 meV) biases. Black and red arrows represent the forward and backward gate sweeps. The inset shows the transport curve of a non-encapsulated device.

(c) Top gate voltage dependence of bias current under vacuum (black line) and ambient conditions (red line) at fixed back gate (0V) and source-drain (100 meV) biases. The inset shows the gate voltage dependence of a non-encapsulated device. All measurements are performed in ~ 4.5 nm thick sample at room temperature.

**Figure 3.** (a-c) 2D color fan plot of bias current as a function of back gate and source-drain biases at fixed top gate voltage values of -4V, 0V, 4V respectively. (d) Back gate voltage dependence of bias current at different top gate voltage values. (e) Source-drain voltage dependence of bias current at different displacement fields. Black, red and blue lines represent $V_{TG}$=-4V, 0V and 4V while $V_{BG}$ is fixed to -30V. Inset shows the renormalized values of Figure 3-(e) for more clarity. All measurements are performed in ~ 4.5 nm thick sample at room temperature. (f) Top and bottom schematics illustrate the band alignments at large negative displacement fields and small displacement fields respectively.

**Figure 4.** (a) Source-drain bias voltage and temperature dependence of bias current at fixed top gate (-4V) and back gate (-70V) voltage. Inset shows the temperature dependence of bias current at different source-bias voltage values. (b) Arrhenius plot for source-drain voltages from 1V (black sphere) to 0.1V (Dark yellow sphere) at fixed back gate (-70V) and top gate (-4V) voltages. (c) The schematic illustrates current injection from the contact to a p-type semiconductor via tunneling ((1) direct or (2) thermally assisted) and (3) thermionic emission mechanisms. (d) Temperature dependence of two-terminal field effect mobility for the ~ 4.5 nm thick sample.

**Figure 5.** Back gate voltage dependence of bias current in a bilayer BP-based device at fixed source-drain voltages (0.1V). Inset shows the source-drain voltage dependence of bias current at fixed back gate voltage values. All measurements are performed at room temperature under vacuum conditions.

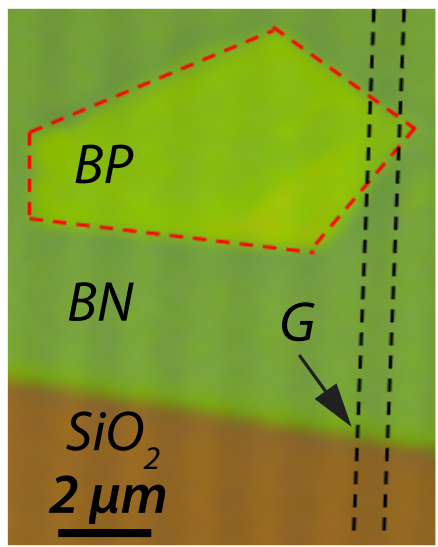
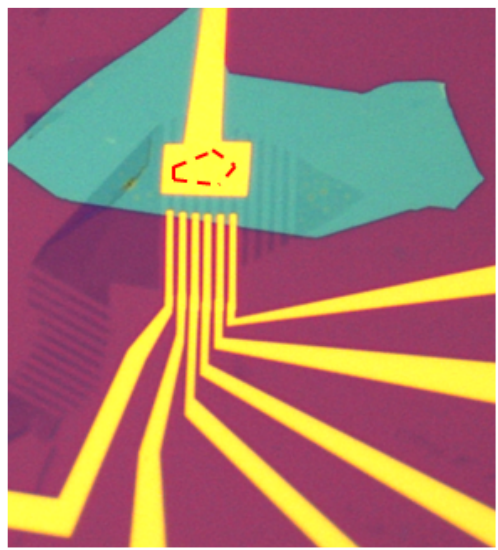
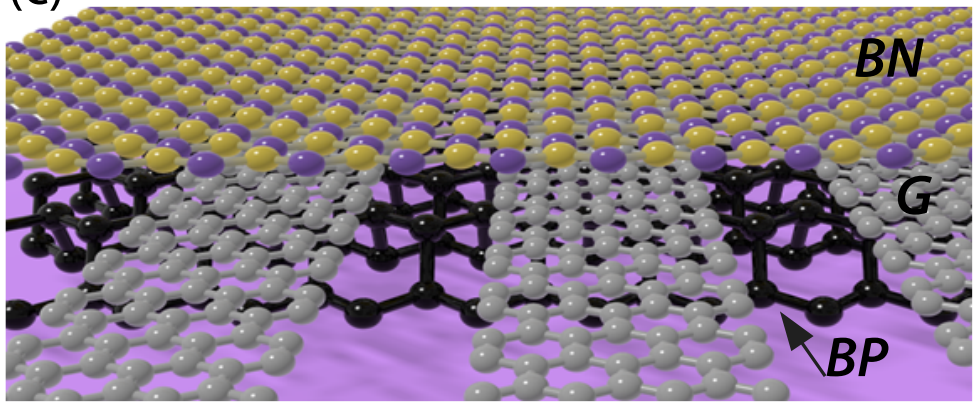
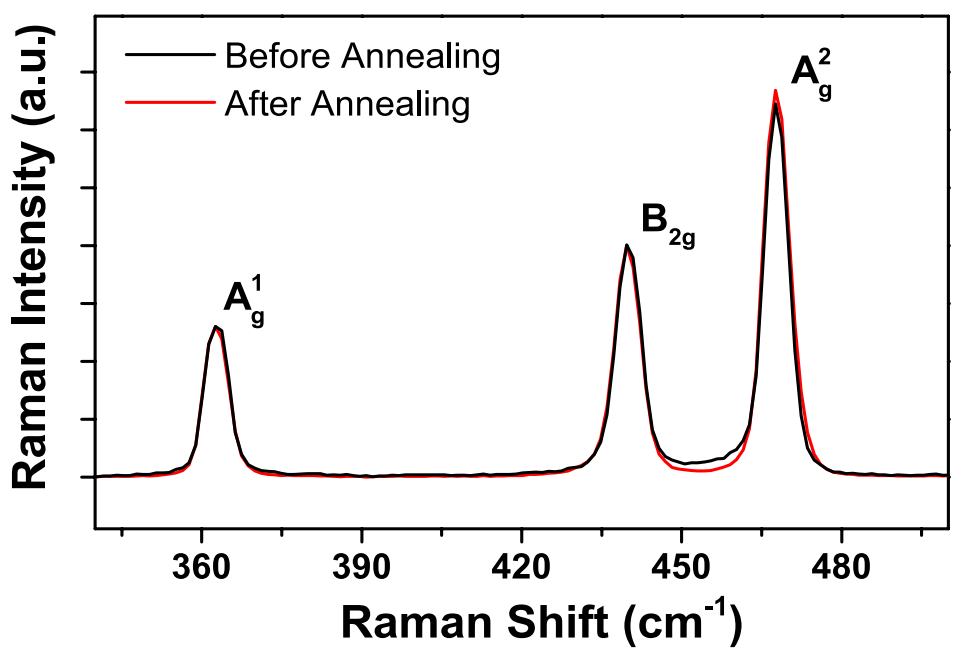

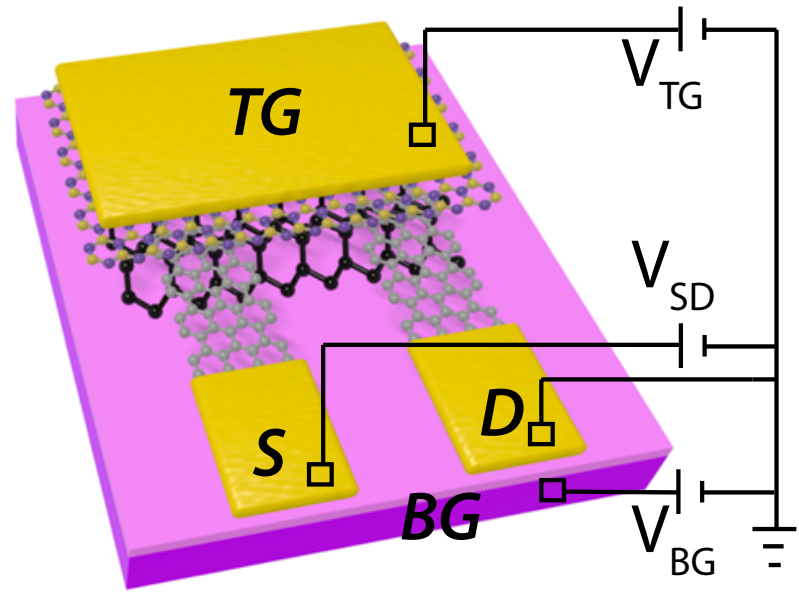 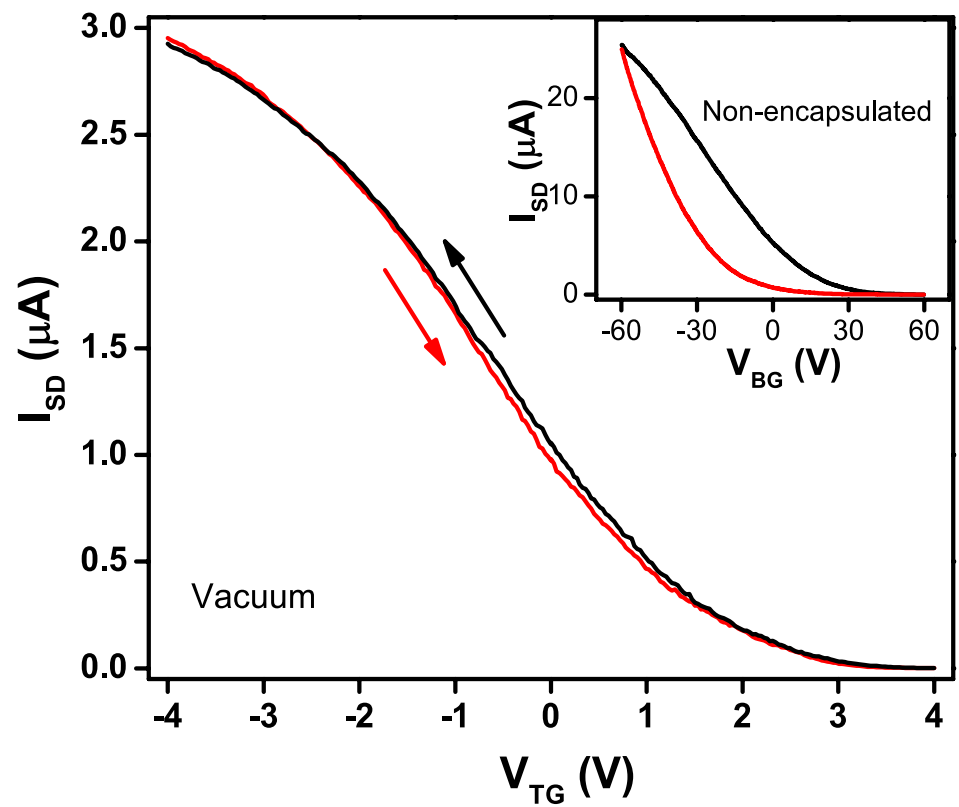 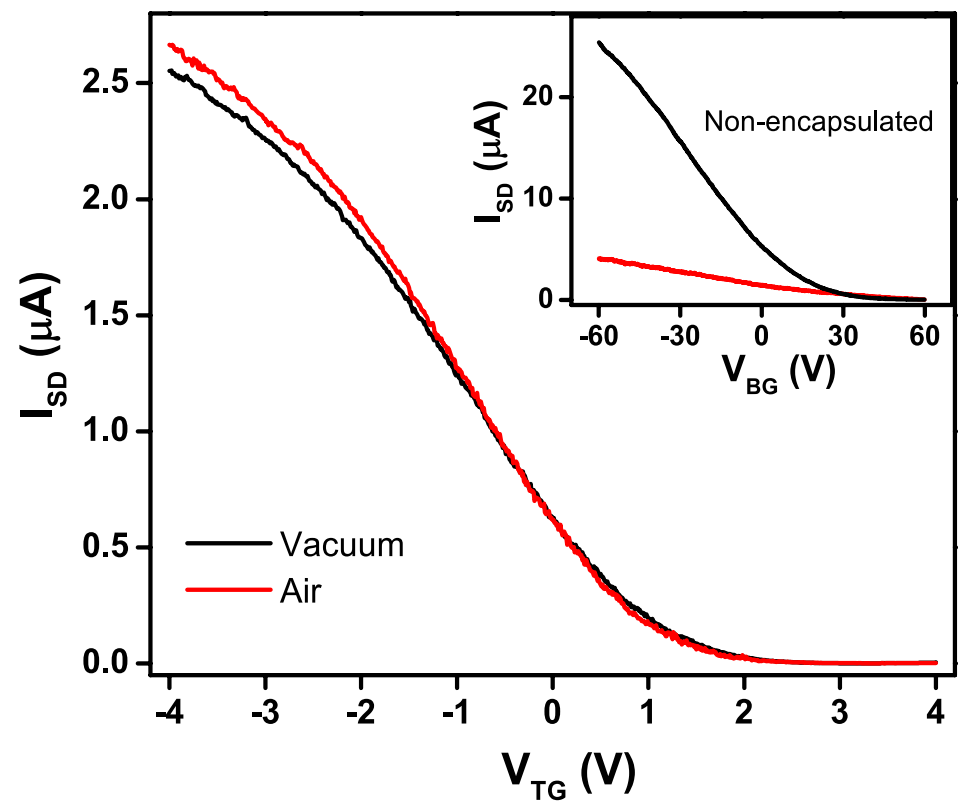

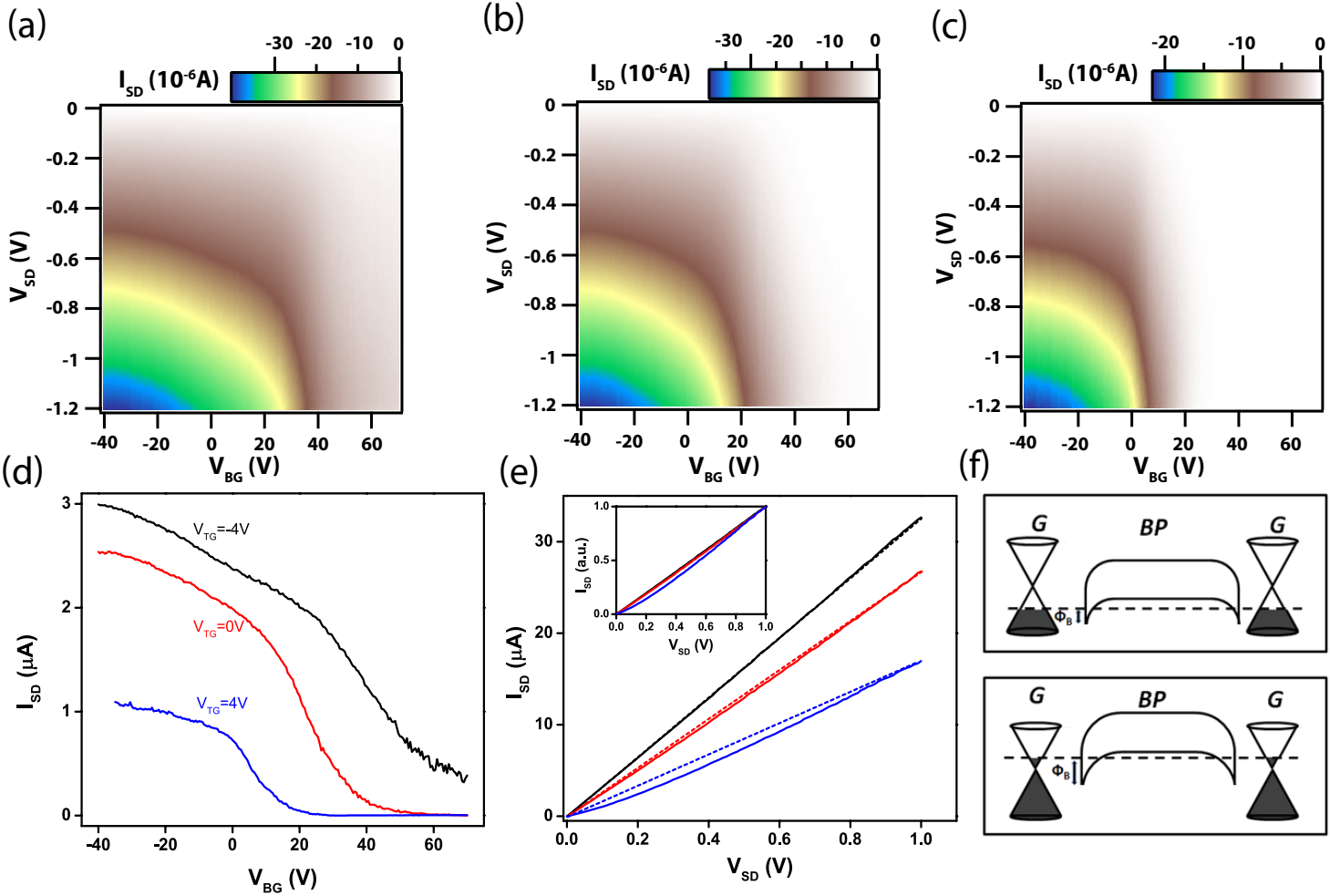

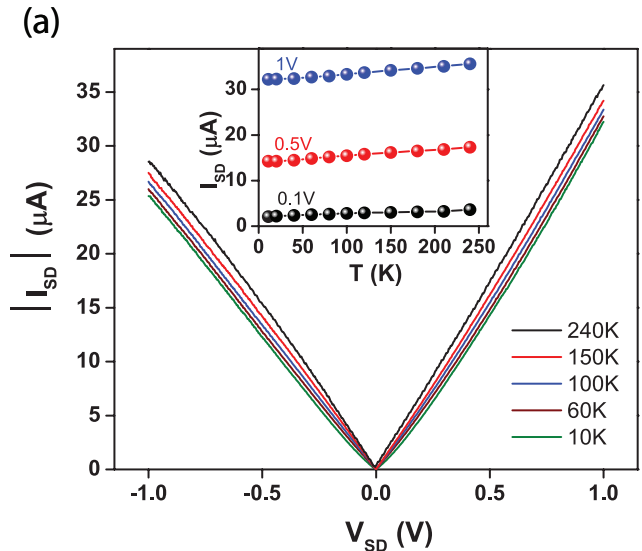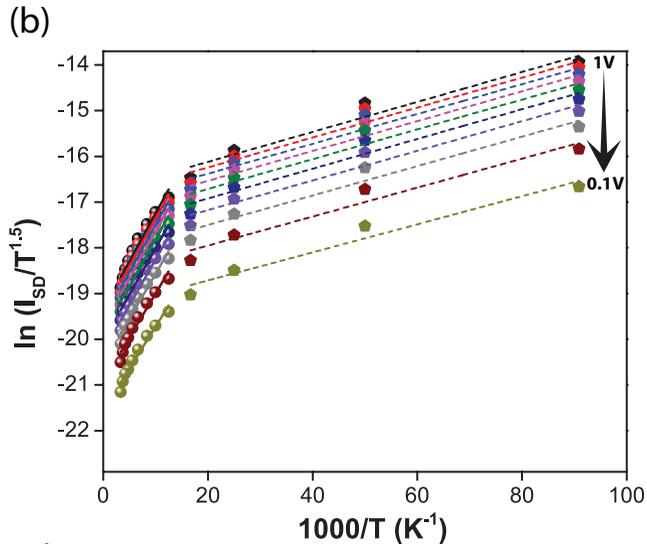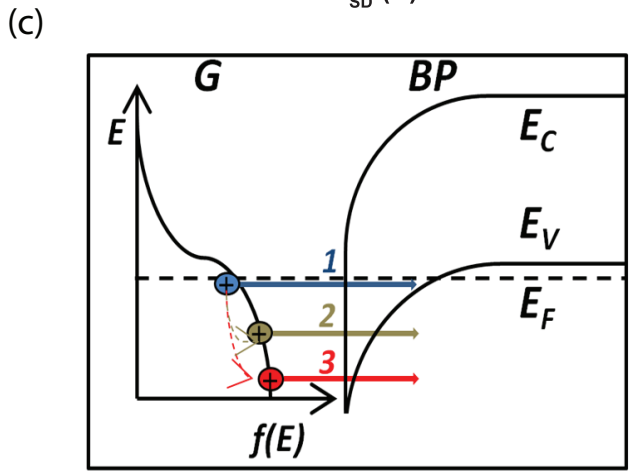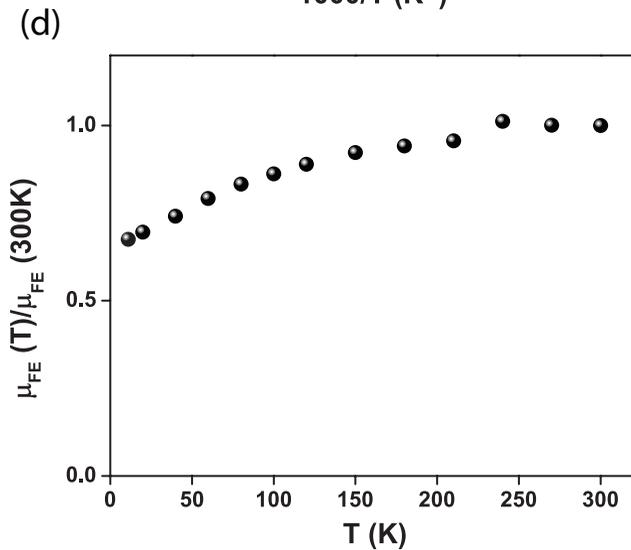

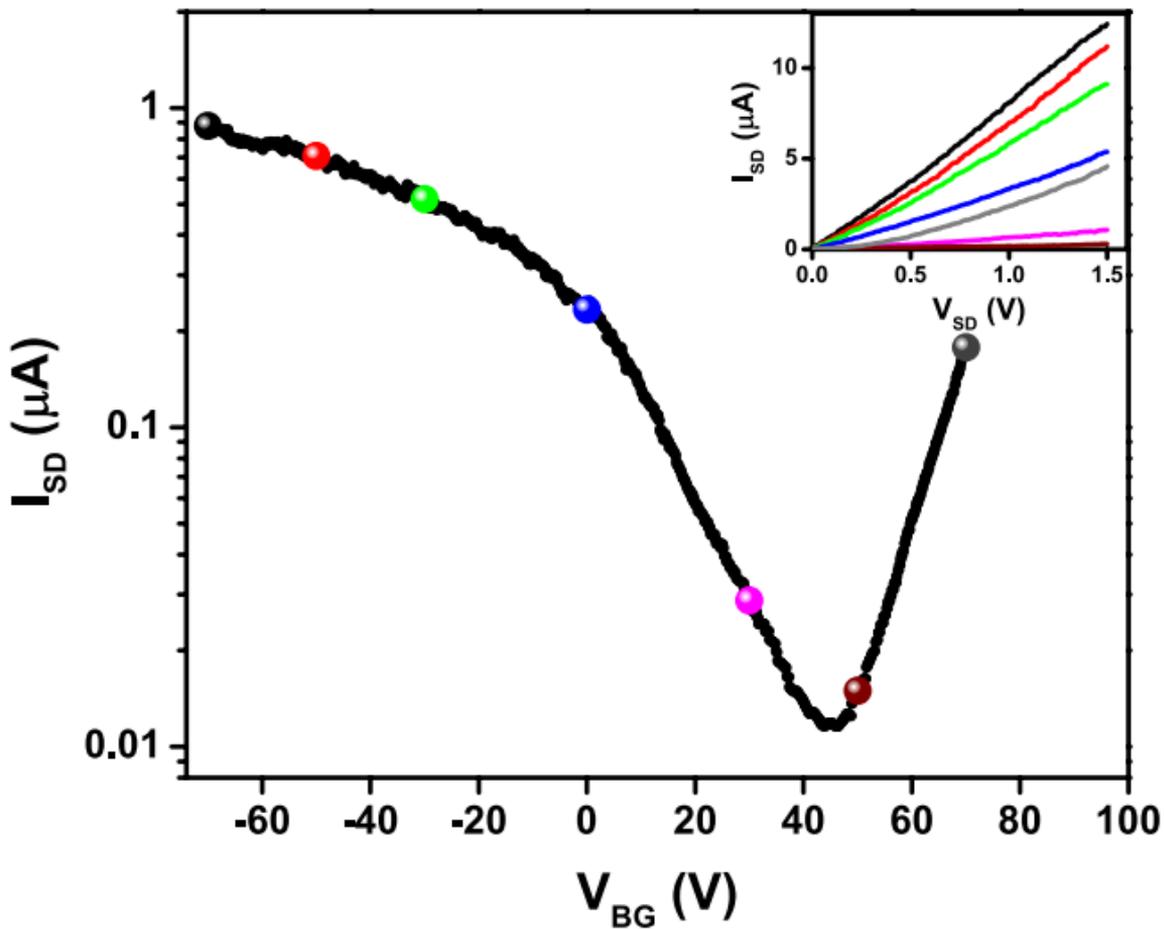

# Supporting Information
# Electrical characterization of fully encapsulated ultra thin black phosphorous-based heterostructures with graphene contacts

Ahmet Avsar[1], Ivan J. Vera-Marun[1,2], Tan Jun You[1], Kenji Watanabe[3], Takashi Taniguchi[3] Antonio H. Castro Neto[1] and Barbaros Özyilmaz[1,4,†]

[1]Graphene Research Center & Department of Physics, 2 Science Drive 3, National University of Singapore, Singapore 17542, Singapore

[2]Physics of Nanodevices, Zernike Institute for Advanced Materials, University of Groningen, Nijenborgh 4, 9747 AG, Groningen, The Netherlands

[3]National Institute for Materials Science, 1-1 Namiki, Tsukuba 305-0044, Japan

[4]Nanocore, 4 Engineering Drive 3, National University of Singapore, Singapore 117576, Singapore

[†]Corresponding author: barbaros@nus.edu.sg,

## Section 1: Fabrication process flow

- Graphene is mechanically exfoliated onto Si/SiO$_2$ (300nm) substrate and etched into stripes to serve as source and drain electrodes.

- Boron Nitride (BN) is exfoliated onto PMMA/PMGI/Si/SiO$_2$ stack. BN is isolated from the wafer by removing the PMGI resist layer similarly discussed in ref.1.

- BN is transferred partially onto graphene stripes with a dry transfer method by using a home-made transfer stage under ambient conditions (Figure S1-a). The heterostructure is annealed at 340 C under Ar/H$_2$ gas environment for 6 hours. This process removes the possible polymer residues and small bubbles formed at the graphene/BN interface.

- A thin layer of PMMA is span onto the heterostructure and the resulting G/BN/PMMA stack is isolated from Si/SiO$_2$ wafer with a KOH etching step (Figure S1-b). The stack is cleaned carefully with DI water.

- BP crystal is exfoliated just before the final transfer process. As shown in Figure S1-c, G/BN/PMMA stack is transferred onto BP crystal in Argon filled glove box (negligible O$_2$ and H$_2$O concentration). It is ensured that BN fully encapsulated BP thereby the surface of BP never be exposed to air.

- Cr/Au (2 nm/80 nm) source-drain and top gate contacts are deposited onto non-encapsulated region of graphene and on top of BN crystal, respectively (Figure S1-e).

- The final device is annealed at 340 C under high vacuum conditions to enhance the adhesion between these 2D crystals. The optical and AFM images of final device are shown in Figure S1-f.

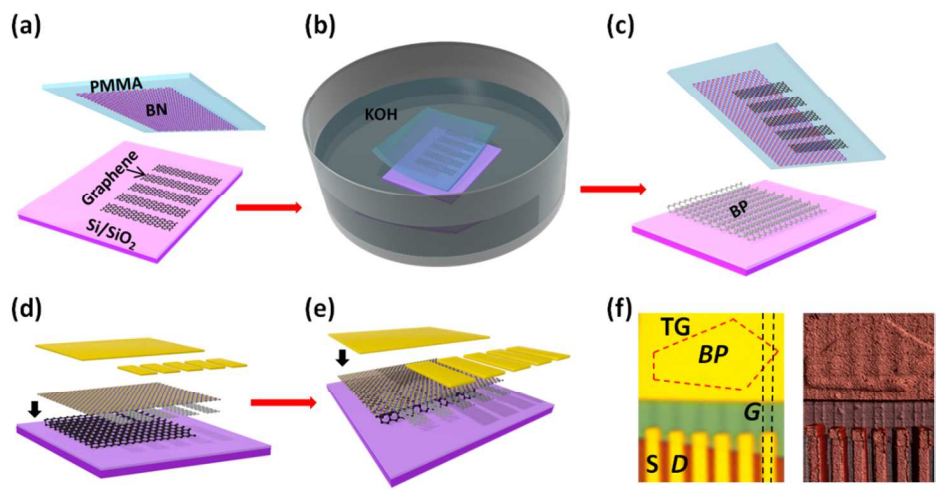

Figure S1: The schematics of device fabrication process for encapsulated black phosphorous device. Process shown at (c) is done under Argon environment. (f) shows the optical and AFM images of the completed device.

**Section 2: Effect of annealing on the stability of non-encapsulated black phosphorous crystals**

The annealing treatment in our device process is necessary to further improve the performance of the BP-based heterostructures. The annealing treatment is done at 340 C for 3 hours under high vacuum conditions. The encapsulation process with BN crystal physically protects the BP, without showing any signature of the degradation. However, the non-encapsulated BP crystals are very sensitive to the environment. Figure S2-a and S2-b show the optical images of a thin BP crystal before and after annealing process. As clearly can be seen, the non-encapsulated thin flake is fully degraded during the annealing process (Figure S2-c).

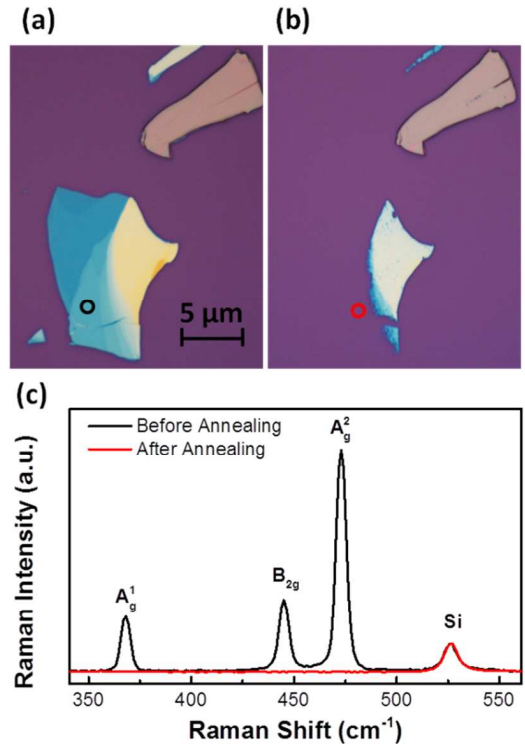

Figure S2: (a-b) Optical images of BP crystal before and after the annealing treatment. Black and red circles represent the Raman spectroscopy locations. (c) Raman spectrum of BP crystal before and after annealing treatment.

## Section 3: Hysteresis in non-encapsulated devices

The rapid degradation of non-encapsulated BP crystals at ambient conditions may affect the transport properties of devices substantially. We measured few non-encapsulated devices under both ambient and vacuum conditions at room temperature. Figure S3-a shows the $V_{BG}$ dependence of $I_{SD}$ at fixed $V_{SD}$ (100 meV). The hysteresis increases from ~ 28V to ~ 60V upon exposing the sample to the air. The hysteresis also increases as the $V_{BG}$ range increases. Such trend is caused by a charge trapping mechanism associated with the sensitivity of BP to adsorbents present at air and host substrate[1]. Figure S3-c and S3-d show the hysteresis and current modulation ratio in different samples measured under both vacuum and ambient conditions. The thicknesses of flakes in sample 1, 2, 3 and 4 are 5.5 nm, 6 nm 7 nm and 12 nm, respectively. While we could not make a clear relation between thickness of the crystal and hysteresis, all samples show poorer device performance upon exposing to air. This suggests that passivation of BP devices is critical to obtain realiable and repeatable measurement.

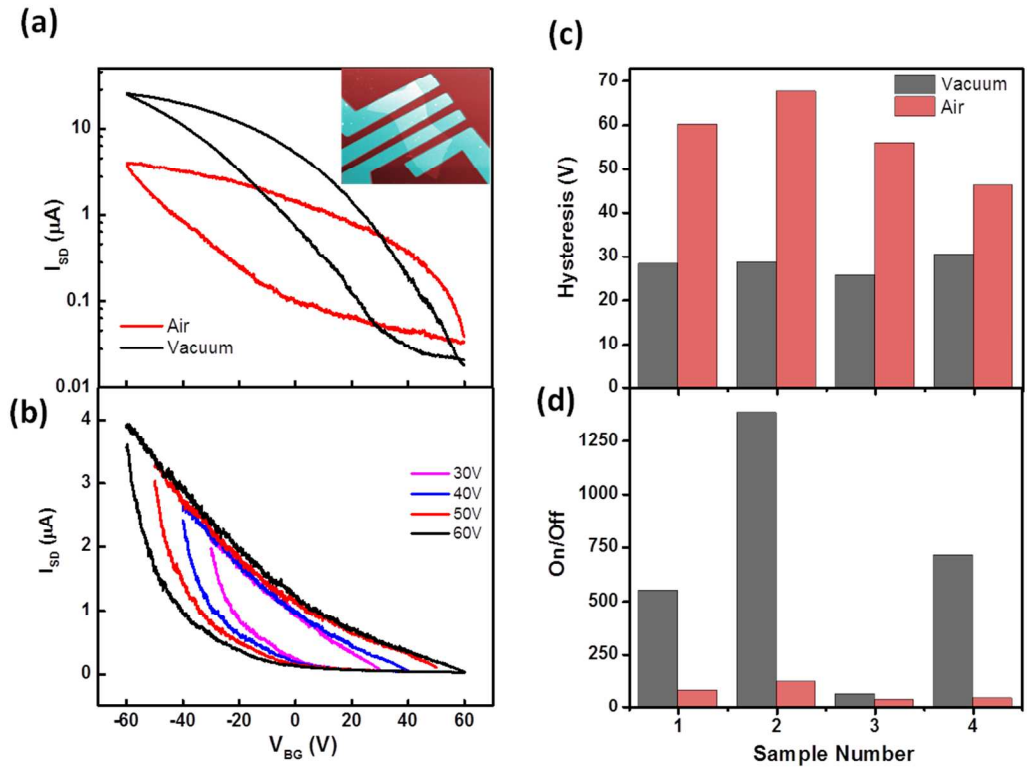

Figure S3: (a) Hysteresis measurement under both vacuum and ambient conditions for a non-encapsulated device (thickness ~ 5.5nm). Inset is the AFM image of the device. Color height is 0-50 nm. (b) Hysteresis measurement under ambient conditions for different $V_{BG}$ ranges. (c-d) Plots show change in hysteresis and on/off ratios under both ambient and vacuum conditions for different thicknesses samples.

**Section 4: Charge transport in a non-encapsulated device at low temperature**

The thermal fluctuation at room temperature is in the order of Schottky barrier height for metal contacted devices[2]. This energy allows the charge carriers to overcome the barrier and results in linear I-V relation at room temperature[3,4]. The linear I-V relation at RT can mislead to believe that an Ohmic contact is achieved. Here we study the transport properties of a metal contacted, non-encapsulated device at 4.2 K. As can be seen in Figure S4-b, the I-V is non-linear at 4.2 K even at low bias voltage. This result is consistent with the work of Kamalakar et al. where they observe a non-linear I-V behavior even at 200 K in metal contacted BP FETs[2]. As discussed earlier, the charge injection in these metal contacted devices is primarily dominated by thermionic emission mechanism. We note that our graphene contacted, encapsulated BP devices have linear I-V even at low temperature and high bias regime.

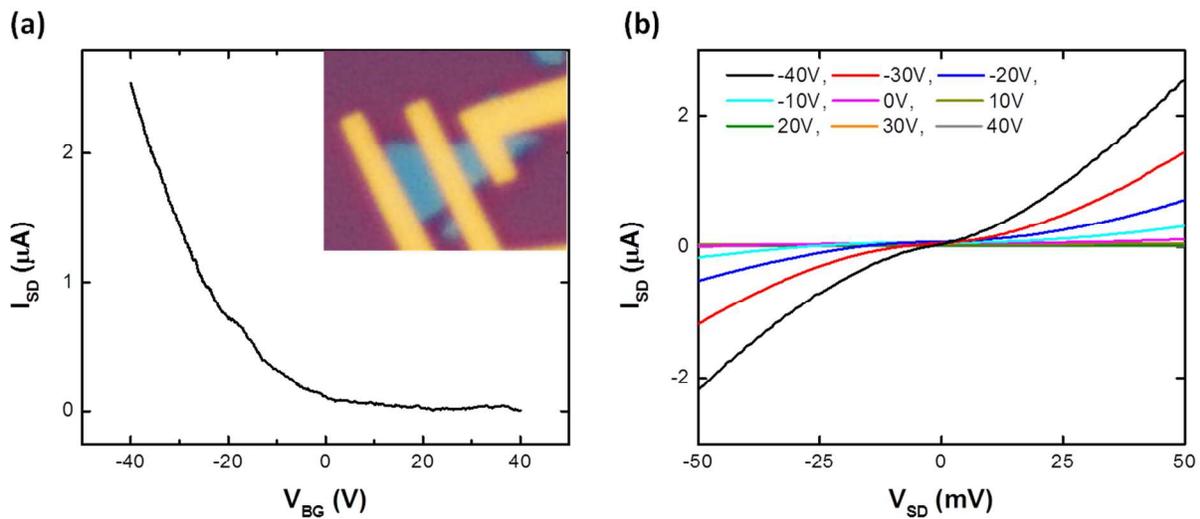

Figure S4: (a) $V_{BG}$ dependence of $I_{SD}$ at fixed $V_{SD}$ (100 meV). The inset shows the optical image of the measured device. (b) $V_{SD}$ dependence of $I_{SD}$ at various $V_{BG}$ values.

**Section 5: Robust transport in encapsulated device**

Figure S5-a and S5-b show the transport characteristic of an encapsulated device two months after its fabrication was completed. The device shows nearly hysteresis free transport, indicating that BN encapsulation allows realiable operations over time without any signature of the degradation. The $I_{SD}$ is recorded as a function of both $V_{BG}$ and $V_{TG}$ to prove that hysteresis free transport is observed with gating through both the shown hysteresis free transport in main text is not due to thin top gate dielectric.

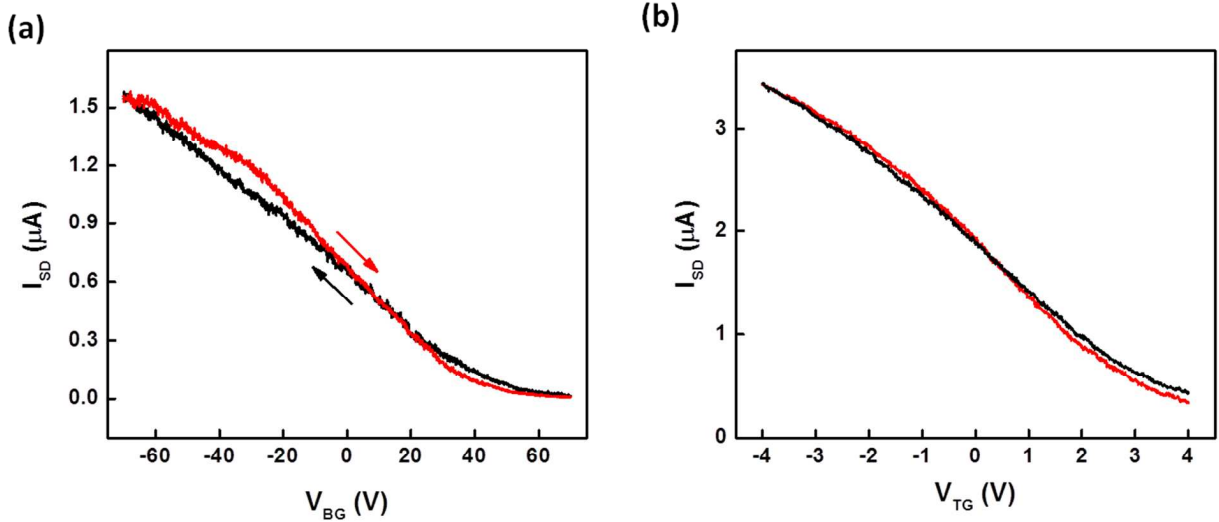

Figure S5: (a) Room temperature $V_{BG}$ dependence of $I_{SD}$ at fixed $V_{TG}$ (0 V) and $V_{SD}$ (100 meV). (b) $V_{TG}$ dependence of $I_{SD}$ at fixed $V_{BG}$ (-70 V) and $V_{SD}$ (100 meV). The measurement is performed two months after the device fabrication was completed.

## Section 6: Additional hysteresis data

Figure S6 shows the room temperature $V_{BG}$ dependence of $I_{SD}$ at fixed $V_{TG}$ (0 V) and $V_{SD}$ (100 meV) for a graphene contacted encapsulated BP device. The thickness of BP is ~ 6.5 nm. Similar to the data shown in manuscript, we observe nearly hysteresis free transport.

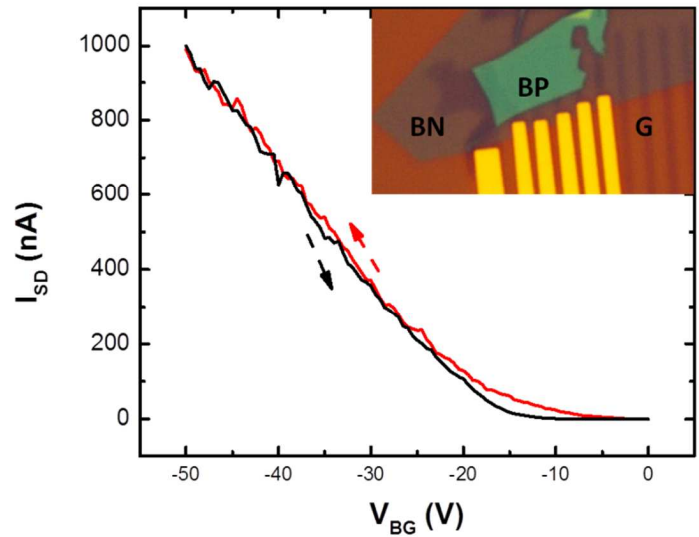

Figure S6: (a) Room temperature $V_{BG}$ dependence of $I_{SD}$ at fixed $V_{TG}$ (0 V) and $V_{SD}$ (100 meV). Red and black arrows represent the $V_{BG}$ sweep directions. The inset shows the optical image of the measured device.

## Section 7: Encapsulated devices contacted with Ti/Au contacts

As discussed in main text, the flexibility of monolayer graphene allows the BN passivation layer to seal BP very tightly. The graphene work function tunability results in ideal contact behavior. In order to test the performance of commonly used Ti/Au contacts in encapsulated BP device, we fabricated a Ti/Au (5nm/80nm) contacted BP-FET. A thin layer of BP (~6.5 nm) is exfoliated on $SiO_2$ substrate at glove box. A PMMA layer is span in glove box on the crystal for contact patterning and also to protect the sample from air during process. The surface of BP is exposed to the air at the contact regions after the resist development just before the metallization process. The lift off process is done in glove box followed by the transfer of a BN layer on top of the device. Figure S7-a shows the forward and backward scans of $V_{BG}$ while $I_{SD}$ is recorded. High hysteresis is observed in this device even it is encapsulated. This hysteresis is caused due to surface degradation at exposed contact area during the fabrication process. Non-flexible, thick Ti/Au contacts do not allow BN to seal BP perfectly, and samples degrades over time due to diffusion of gaseous from environment. Last but not least important, Ti/Au contacts do not show a linear relation between $I_{SD}$ and $V_{SD}$ even at low biases (Figure S7-b). All these results show the signifance of graphene contacts in the presented device architecture.

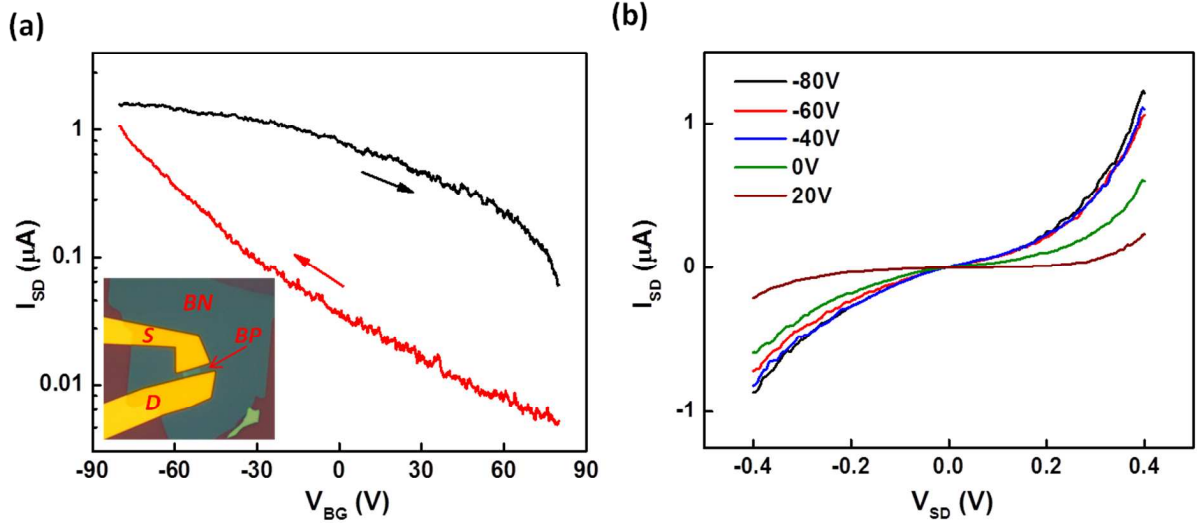

Figure S7: (a) Room temperature $V_{BG}$ dependence of $I_{SD}$ at fixed $V_{SD}$ (100 meV) at vacuum conditions. Red and black arrows show the forward and backward sweep directions respectively. Inset: Optical image of Ti/Au contacted encapsulated BP device. (b) $V_{sd}$ dependence of $I_{SD}$ at fixed $V_{BG}$ values.

**Section 8: Discussion on the contribution of thermionic emission for the charge injection**

Figure S8-a shows the Arrehenius plots recorded for $V_{BG}$= 50V while $V_{TG}$ is kept fixed at -4V. Unlike previous reports which were utilizing Ti/Co metals as contact material[2], we observe negative slope for the Arrehenius plots at graphene contacted devices. The negative slope indicates that extracted Schottky barrier heights with the thermionic model are negative which clearly demonstrate that thermionic emission is not dominant in our graphene contacted devices (Figure S8-b). The observation of such extremely reduced Schottky barrier heights even at large $V_{BG}$ values (displacement field is low) makes graphene contacts ideal for BP based FETs.

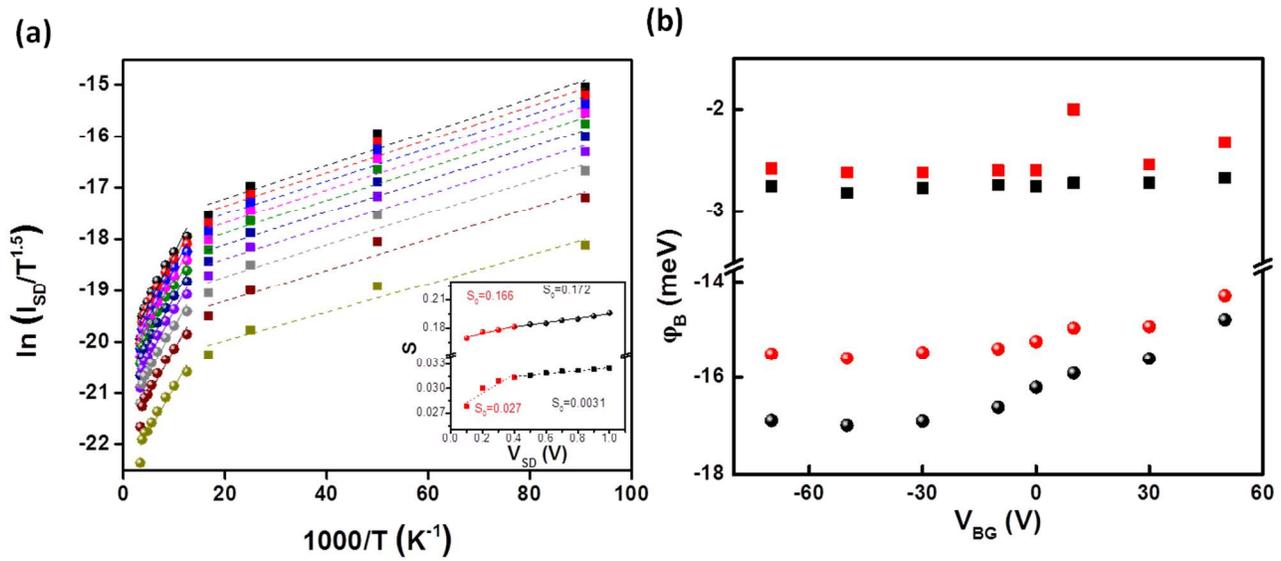

Figure S8: (a) Arrhenius plots of $V_{SD}$ varying from 1V (Black) to 0.1V (Dark yellow) at $V_{BG}$= 50 V and $V_{TG}$= -4 V respectively. (b) $V_{BG}$ dependence of Schottky barrier height extracted from the Arrhenius plots at different temperature and bias ranges. Red and black squares represent the Schottky barrier heights at low (0.1V-0.4V) and high (0.5V-1V) $V_{SD}$ values at low temperature range (11K-60K) respectively. Red and black spheres represent the Schottky barrier heights at low (0.1V-0.4V) and high (0.5V-1V) $V_{SD}$ values at high temperature range (80K-300K) respectively.

Figure S9 shows the temperature and $V_{BG}$ dependence of the $I_{SD}$-$V_{SD}$ curve at fixed $V_{TG}$ (-4V) for the ~ 4.5 nm thick encapsulated BP device, shown in manuscript. We note that temperature dependence of $I_{SD}$ even at high positive $V_{BG}$ is very weak and the charge injection is not dominated by the thermionic emission mechanism as discussed before in Figure S8.

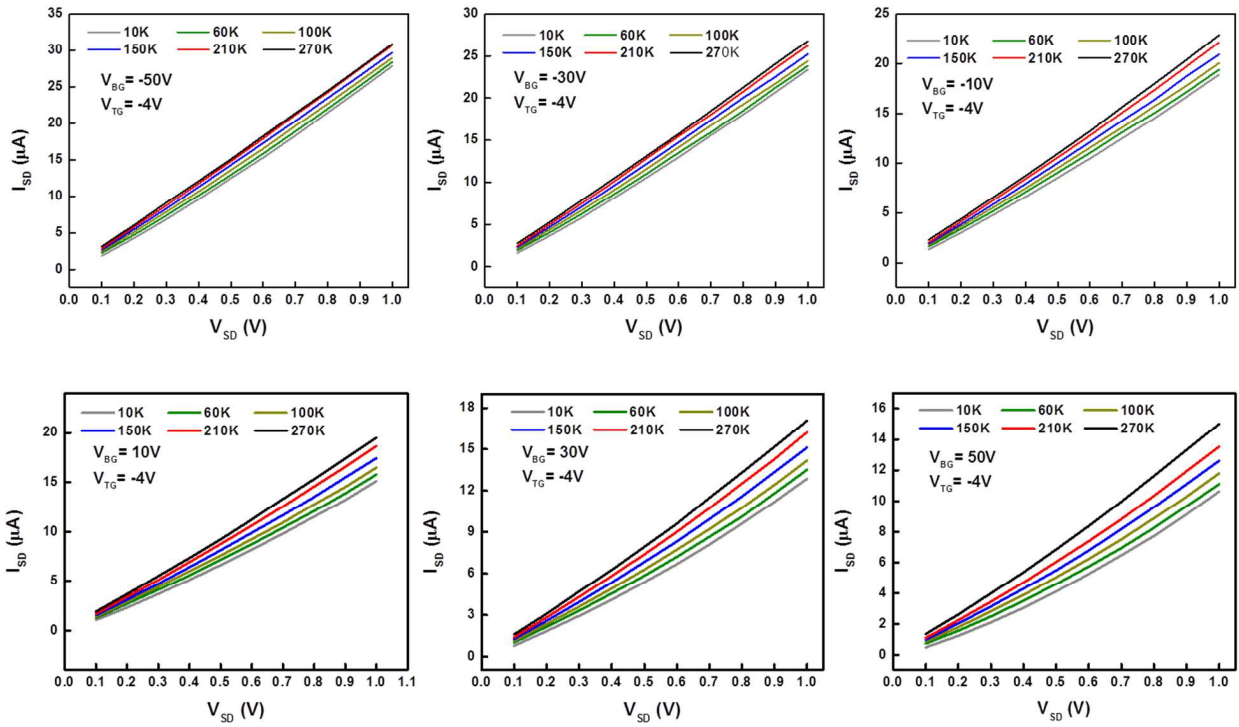

Figure S9: $V_{SD}$ dependence of $I_{SD}$ at different $V_{BG}$ and temperature values.

## Section 9: Work function tunability of graphene electrodes

As discussed in Figure 3 d-f, the application of vertical electric field tunes the work function of graphene[5] and results in more effective charge injection into black phosphorous. This leads to a linear I-V at high negative displacement fields (both $V_{BG}$ and $V_{TG}$ negative). Here, we show the both $V_{BG}$ and $V_{TG}$ dependence of $I_{SD}$ at room temperature. Similar to Figure 3-e, I-V deviates from perfect linearity as $V_{TG}$ is swept from -4V to 4V.

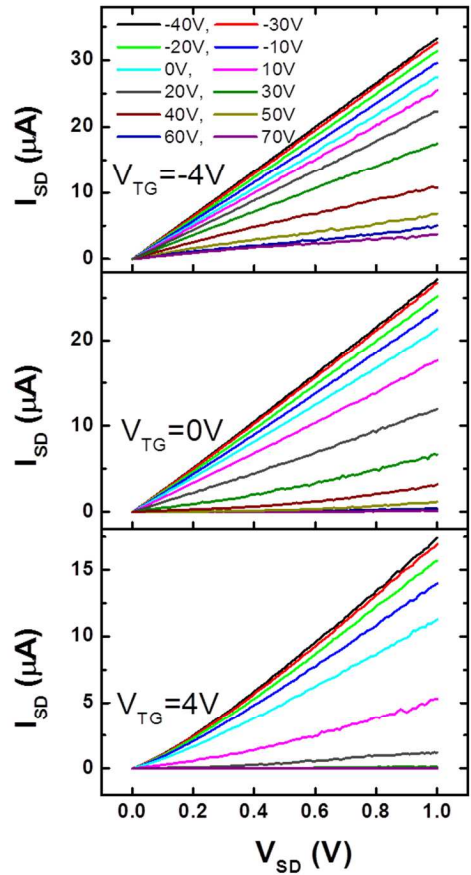

Figure S10: $V_{SD}$ dependence of $I_{SD}$ at different $V_{BG}$ and $V_{TG}$ at room temperature.

**Section 10: Current modulation and subthreshold swing in encapsulated device**

Figure S11-a shows the $V_{TG}$ and $V_{BG}$ dependence of $I_{SD}$ at room temperature for the thicker device shown in manuscript. The device does not show the off state in $V_{BG}$ sweep if $V_{TG}$ > -0.5V. Figure S11-b shows the semi-log scale plot of $I_{SD}$ vs $V_{BG}$ at different $V_{TG}$ values. The on/off ratio for this device exceeds $10^4$ for hole charge carriers. Subthreshold swing is only ~ 2 V per decade. We note that the on/off ratio and subthreshold swing values are sample dependent, even shows variations at the different junctions in the same crystal. Figure S11-c shows the same plot for another encapsulated BP device. On/off ratio is ~ 3.5 x $10^3$ and subthreshold swing is ~ 4.5 V per decade. We note that the observed subthreshold swing values are slightly improved compared to the previous reports with similar thick $SiO_2$ dielectrics[2,4]. Compared to $MoS_2$ FETs which have near-ideal subthreshold swings of ~ 70 mV/dec[6], the larger values in our devices are mainly due to the thick $SiO_2$ dielectric used in our experiment. Thinner substrates with higher dielectric constants can significantly improve the subthreshold swing of BP.

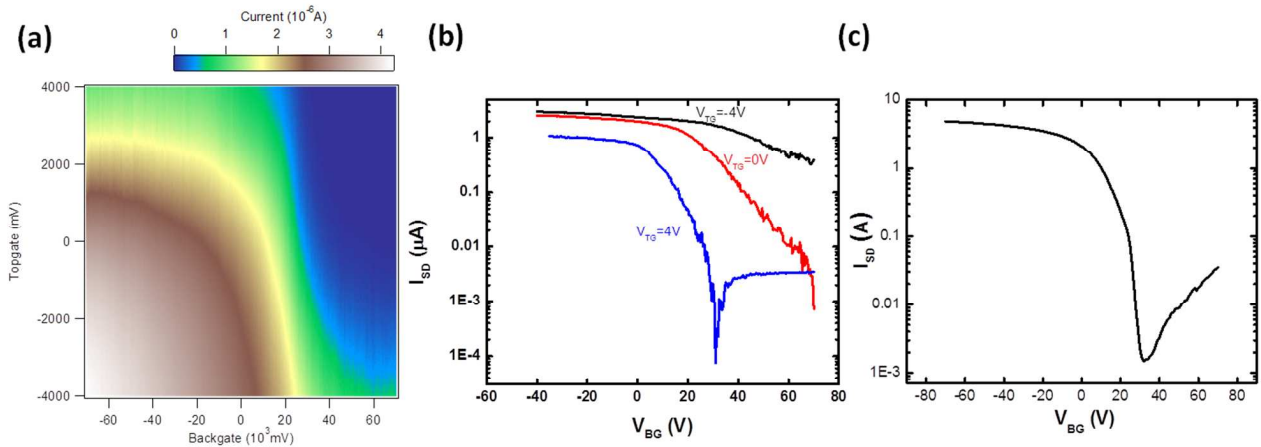

Figure S11: (a) $V_{BG}$ and $V_{TG}$ dependence of $I_{SD}$ at fixed $V_{SD}$ (0.1V) at room temperature. (b) Semi-log scale plot of $I_{SD}$ as a function of $V_{BG}$ at different $V_{TG}$ values.

**Section 11: Temperature dependent gate sweep of bias current in encapsulated device**

Figure S12 shows the $V_{BG}$ dependence of $I_{SD}$ at different temperature values. $V_{TG}$ (-4V) and $V_{SD}$ (0.1V) are fixed during the measurement. The sample shows very weak temperature dependence because of the tunneling mechanism dominated charge injection rather then thermionic emission mechanism. The mobility of the device is measured at the slope where we observe very weak temperature dependence.

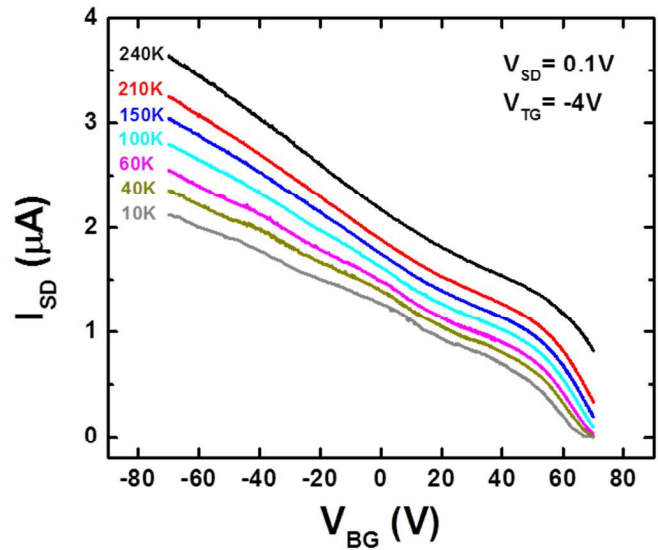

Figure S12: $V_{BG}$ and temperature dependence of $I_{SD}$ at fixed $V_{TG}$ and $V_{SD}$ values